\documentclass[final,5p,times,twocolumn]{elsarticle}
 \biboptions{comma,sort&compress}
\usepackage{graphicx}
\usepackage{here}
\usepackage[dvips]{epsfig}
\def\beq{\begin{equation}}
\def\eeq{\end{equation}}
\def\bea{\begin{eqnarray}}
\def\eea{\end{eqnarray}}
\def\bq{\begin{quote}}
\def\eq{\end{quote}}

\def\nnb{\nonumber}
\def\ga{\left(}
\def\dr{\right)}

\def\nnb{\nonumber}
\def\la{\langle}
\def\ra{\rangle}
\def\nin{\noindent}
\def\ba{\vspace*{-0.2cm}\begin{array}}
\def\ea{\end{array}\vspace*{-0.2cm}}

\def\als{\alpha_s}

\def\gg2{ \la\alpha_s G^2 \ra}
\def\gg3{g^3f_{abc}\la G^aG^bG^c \ra}
\def\ggg4{\la\als^2G^4\ra}

\def\beq{\begin{equation}}
\def\enq{\end{equation}}
\def\beqa{\begin{eqnarray}}
\def\enqa{\end{eqnarray}}
\def\nnb{\nonumber}

\def\MeV{\nobreak\,\mbox{MeV}}
\def\GeV{\nobreak\,\mbox{GeV}}

\def\qq{\lag\bar{q}q\rag}

\def\ss{\lag\bar{s}s\rag}
\def\mix{\lag\bar{q}g\si.Gq\rag}

\def\Gd{\lag g^2G^2\rag}
\def\G3{\lag g^3G^3\rag}
\def\kab{\left[(\al+\be)m_c^2-\al\be s\right]}

\def\rh{\rho}
\def\si{\sigma}

\def\al{\alpha}
\def\be{\beta}
\def\alma{\alpha_{max}}
\def\almi{\alpha_{min}}
\def\bemi{\beta_{min}}
\def\lb{\label}
\def\nn{\nonumber}

\newcommand{\rag}{\rangle}
\newcommand{\lag}{\langle}

\journal{Physics Letters B}

\begin{document}

\begin{frontmatter}

\title{Relation between $T_{cc,bb}$ and $X_{c,b}$ from QCD}
 \author[label1]{J.M.  Dias}
\ead{jdias@if.usp.br}
\address[label1]{Instituto de F\'{\i}sica, Universidade de S\~{a}o Paulo, 
C.P. 66318, 05389-970 S\~{a}o Paulo, SP, Brazil}
%
 \author[label2]{S. Narison}
\ead{snarison@yahoo.fr}
\address[label2]{Laboratoire
Particules et Univers de Montpellier, CNRS-IN2P3  , 
\\
Case 070, Place Eug\`ene
Bataillon, 34095 - Montpellier Cedex 05, France.}

 \author[label1]{F. S.  Navarra}
\ead{navarra@if.usp.br}
%
\author[label1]{M. Nielsen\corref{cor1}}
\cortext[cor1]{Corresponding author}
\ead{mnielsen@if.usp.br}
\author[label3]{J.-M. Richard}
\ead{j-m.richard@ipnl.in2p3.fr}
\address[label3]{Universit\'e de Lyon et Institut de Physique Nucl\'eaire de Lyon,
IN2P3-CNRS-UCBL \\
4, rue Enrico Fermi, F-69622 Villeurbanne, France}

\begin{abstract}
\nin
We have studied, using double ratio of QCD (spectral) sum rules, the ratio between the
masses of $T_{cc}$ and $X(3872)$ assuming that they are respectively described  by the $D-{D}^*$ and $D-\bar{D}^*$ molecular
currents. We found (within our approximation) that 
the masses of these two states are almost degenerate. 
Since the pion exchange interaction between
these mesons is exactly the same, we conclude that if the observed $X(3872)$ 
meson
is a $D\bar{D}^*+c.c.$ molecule, then the $DD^*$ molecule should also exist with
approximately the same mass. An extension of the analysis to the $b$-quark case 
leads to the same
conclusion. We also study the SU(3) breakings for the $T^s_{QQ}/T_{QQ}$  mass ratios. 
Motivated by the recent Belle observation of two $Z_b$ states, we revise our determination
of $X_b$ by combining results from exponential and FESR sum rules.  
\end{abstract}
\end{frontmatter}
\section{Introduction}
The existence of  exotic hadrons is a long-standing problem. By exotic we mean a state whose quantum numbers and 
main properties cannot be explained by a simple quark-antiquark or three-quark 
configuration. 

The $X(3872)$ resonance (assumed to be an $1^{++}$ axial vector meson) has, indeed, stimulated {\it many activities in the physics of hadrons}. 
It was  discovered by BELLE in $B$-decays \cite{BELLE}, and confirmed by BABAR 
\cite{BABAR}, CDF \cite{CDF} and D0 \cite{D0}. It is rather narrow, with a width $\leq$ 2.3 MeV.
Its most popular picture  which consists of a molecular configuration, $D\bar{D}^*+\bar{D}D^*$, with $J^{PC}=1^{++}$, 
has been attributed to the narrow ($\leq$ 2.3 MeV width) $X(3872)$ 
\footnote{For references about some other possible interpretations of the $X(3872)$
see, e.g., \cite{rev}.}.

The case of the four-quark state $(QQ\bar{u}\bar{d})$
with quantum numbers $I=0, ~J=1$ and $P=+1$ which, following ref.\cite{ros},
we call $T_{QQ}$, is especially interesting. As already noted previously
\cite{ros,zsgr},  the $T_{bb}$ or $T_{cc}$ states with $J^P=1^+$ cannot split into a 
pair of  two $\bar B$ or two $D$ mesons which is retricted to $J^P=0^+, 2^+, \ldots$. If 
their masses are below the $\bar B\bar B^*$ or $DD\pi$ thresholds, these decays are also 
forbidden. As a result, $T_{QQ}$ becomes stable with respect to strong interaction, and 
must decay radiatively, or even weakly if the mass becomes lower than the threshold 
made of two pseudoscalar mesons.

\section{$T_{QQ}$ from potential models}
In constituent models with a flavor-independent central potential, the stability of $(QQ\bar q\bar q)$ 
configurations comes from a favorable effect when the charge-conjugation symmetry is broken, as noted many years ago \cite{art}.
This is the same mechanism by which, in QED, the loosely bound positronium molecule evolves into the very stable hydrogen molecule.

It is worth noting that in the
large $m_Q$ limit, the light degrees of freedom cannot resolve the closely
bound $QQ$ system. This results in bound states similar to the
$\bar{\Lambda}_Q$ states, with $QQ$ playing the role of the heavy antiquark~\cite{chow}. 

The $(QQ\bar q\bar q)$ states have been studied using a variety of simple or elaborated 
potential models \cite{art,zsgr,vvt,sem,brst}. The corresponding four-body problem is 
very delicate.  For instance, an expansion on harmonic-oscillator states was used in 
\cite{sem}. It is efficient for deep binding but converges very slowly for weak binding. 
If truncated, this expansion may fail to demonstrate stability with potentials that do 
bind, because it lacks explicit $(Q\bar q)-(Q\bar q)$ components, which are important 
near threshold \cite{zsgr}, and are included in the Gaussian expansion sketched in 
\cite{brst} and systematically developed in \cite{ros}. See, also, 
Ref.~\cite{Vijande:2009kj} for a discussion about the four-quark problem.  All authors 
agree that such states become bound when the quark over the antiquark mass ratio becomes 
sufficiently large. Detailed four-body calculations, using a pairwise central potential 
supplemented by a chromomagnetic interaction, indicate that $T_{bb}$ is rather well bound,
 and $T_{cc}$ possibly bound by a few MeV below $DD^*$. For instance, the prediction of 
Ref.~\cite{ros} is, in units of MeV:
\beq
 M_{T_{cc}}=3876 \sim 3905~,~~~
M_{T_{bb}}=10519 \sim 10651~.
\lb{eq:potential}
\eeq
A non-pairwise confinement has also been considered \cite{capa}, inspired by the large coupling regime of QCD, where it is shown \cite{vvr} that it is more favorable to build stable tetraquarks.  In this improved quark model, as well as in conventional quark models, it is found that $(QQ\bar q \bar q)$ has an energy lower than $(Q\bar Q q\bar q)$. 

Another variant was considered in \cite{psgr}, with a  chiral potential model, 
which includes meson-exchange forces between quarks, instead of  the chromomagnetic interaction.

The existence
of a $D\bar{D}^*+\bar{D}D^*$ molecule was predicted in ref.~\cite{torn2} on the
basis of the pion-exchange dynamics \footnote{ For further references \ on this approach, see e.g. \cite{Swanson}.}. Here, the pion is exchanged between the hadrons, as in the Yukawa theory of nuclear forces. 
The $DD^*\pi$ and $\bar D\bar D{}^*\pi$ vertices are identical, as well as 
the $D^*D^*\pi$ and $\bar D{}^*\bar D{}^*\pi$ ones. There is only an overall change of sign, due to the $G$-parity of the pion.
 Therefore, if the pion-exchange dynamics\footnote{Usually, the $G$ parity rule transforms an attractive potential into a repulsive one. Here, however, it only changes the sign of the transition potential $D\bar D{}^*\to  D^*\bar D$, and thus just a phase in the two-component bound state wave function.}
is able to bind the $D\bar{D}^*+\bar{D}D^*$ molecule, the same is true
for the $D{D}^*$ molecule. The difference between these two states can only
come from the short-range part of the interaction. 
\section{  $T_{QQ}$ from QCD (spectral) sum rules}
The first study of tetraquarks with two heavy quarks within QCD { (spectral)} sum rules (QCDSR) was done in \cite{nnlsr} by using diquark-antidiquark current. This study is revisited and improved in the present paper. Our aim is also to compare in detail the $(QQ\bar q\bar q)$ and $(Q\bar Q q\bar q)$  configurations.
Such a comparison is attempted in  ref.~\cite{zhu}, where the authors study heavy tetraquarks using a crude color-magnetic interaction, with
flavor symmetry breaking corrections. They
assume that the Belle resonance, $X(3872)$, is a $cq\bar{c}\bar{q}$
tetraquark, and use its mass as input to determine the mass of other
tetraquark states. They obtain, in units of MeV:
\beq
M_{T_{cc}}\simeq 3966~,~~~~~~ M_{T_{bb}}\simeq10372~,
\lb{eq:color}
\eeq
in agreement with the previous results in Eq. (\ref{eq:potential}) and the ones from 
QCD (spectral) sum rule, in units of GeV  \cite{nnlsr}:
 \beq
M_{T_{cc}}=4.2\pm 0.2~,~~~~~~~ M_{T_{bb}}= 10.2\pm 0.3~.
\eeq
The short-range part
of the interaction can be tested by the QCD (spectral) sum rules approach
\cite{svz,rry,SNB}. Therefore, in this work, we study the ratio of the masses of the $T_{cc}$ and $X(3872)$ states, by using the double ratios of 
sum rules (DRSR) introduced in \cite{DRSR}, which is widely applied for accurate determinations of the ratios of couplings and masses \cite{SNFBS,SNGh,SNhl,SNme+e-,mnnr,HBARYON,drx} and form factors \cite{SNFORM}. This accuracy is reached due to partial cancellations of the systematics of the method and of the QCD corrections in the DRSR. 
More recently, the DRSR was used to study different possible currents for
the $X(3872)$ \cite{drx}. It was found that (within the accuracy of 
the method) the different  
structures ($\bar 3-3$ and $\bar 6-6$  tetraquarks and $D\bar{D}^*+\bar{D}D^*$ 
molecule) lead to the same prediction for the mass. This result 
could indicate
 that the 
short-range part of the interaction alone may not be sufficient to reveal 
the  nature of the $X(3872)$.
\subsection{Two-point functions and forms of the sum rules}
The two-point functions of the $X(3872)$ 
 (assumed to be an $1^{++}$ axial vector meson) and the $T_{cc}$ (assumed
to be a $J^P=1^+$ state)  is defined as:
\bea
\Pi^{\mu\nu}_i(q)&\equiv&i\int d^4x ~e^{iq.x}\lag 0
|T[j^\mu_{i}(x){j^\nu_{i}}^\dagger(0)]
|0\rag\nnb\\
&=&-\Pi_{1i}(q^2)(g^{\mu\nu}-{q^\mu q^\nu\over q^2})+\Pi_{0i}(q^2){q^\mu
q^\nu\over q^2},
\lb{2po}
\eea
where $i=X,~T_{cc}$. The two invariants, $\Pi_1$ and $\Pi_0$, appearing in 
Eq.~(\ref{2po}) are independent and have respectively the quantum numbers 
of the spin 1 and 0 mesons.

We assume that the  $X(3872)$ and $T_{cc}$ states are described by the molecular currents:
\bea
j_{X}^{\mu}(x) & = & \ga{g\over\Lambda}\dr^2_{\rm eff}{1 \over \sqrt{2}}
\Big{[}
\left(\bar{q}_a(x) \gamma_{5} c_a(x)
\bar{c}_b(x) \gamma^{\mu}  q_b(x)\right)\nnb\\
&&- \left(\bar{q}_a(x) \gamma^{\mu} 
c_a(x)\bar{c}_b(x) \gamma_{5}  q_b(x)\right)
\Big{]}.
\lb{currx}
\eea
and
\bea
j^{\mu}_{T_{cc}}(x)  = \ga{g'\over\Lambda}\dr^2_{\rm eff}
\bigg(\bar{q}_a(x)\gamma_5 c_a(x)\bar{q}_b(x)\gamma^\mu c_b(x)\bigg),
\lb{currt}
\eea
where $a$ and $b$ are color indices. 

In the molecule assignement,  it is assumed that there is an effective local current 
and  the  meson pairs are weakly 
bound by a van der Vaals force in a Fermi-like theory with a strength 
$(g/\Lambda)^2_{\rm eff}$ which has nothing to 
do with the quarks and gluons inside each meson. 

Due to its analyticity, the correlation function, $\Pi_{1i}$ in Eq.~(\ref{2po}), 
can be written in terms of a dispersion relation:
\beq
\Pi_{1i}(q^2)=\int_{4m_c^2}^\infty ds {\rho_i(s)\over s-q^2}+\cdots \;,
\lb{ope}
\enq
where $\pi \rho_i(s)\equiv\mbox{Im}[\Pi_{1i}(s)]$ is the spectral function.  

The sum rule is obtained by evaluating the correlation function in 
Eq.~(\ref{2po}) in two ways: using the operator product expansion (OPE) and
using the information from hadronic phenomenology. In the OPE side 
we  work at leading order of perturbation theory 
in $\alpha_s$, and we consider the contributions from 
condensates up to dimension six. In the  phenomenological side,
the correlation function is estimated by inserting intermediate states 
for the $X$ and $T_{cc}$ states via their couplings $\lambda_i$ to the molecular currents:
\beq\label{eq: decay}
\lag 0 |
j^\mu_{i}|M_i\rag =\lambda_i\epsilon^\mu~. 
\enq
where $M_i\equiv X,~T_{cc}$,  $j_i^\mu$ are the currents in Eqs.~(\ref{currx}) and (\ref{currt}).

Using the  ansatz: ``one resonance" $\oplus$ ``QCD continuum", 
where the QCD
continuum comes from the discontinuity of the QCD diagrams from a continuum 
threshold $t_c$, the phenomenological side of Eq.~(\ref{2po}) can be written as:
\beq
\Pi_{\mu\nu}^{phen}(q^2)={\lambda_i^2\over
M_i^2-q^2}\left(-g_{\mu\nu}+ {q_\mu q_\nu\over M_i^2}\right)
+\cdots\;, \lb{phe} \enq
where the Lorentz structure $g_{\mu\nu}$ projects out the $1^{+}$ state.
The dots
denote higher axial-vector resonance contributions that will be
parametrized, as usual, by the QCD continuum. 
After making an inverse-Laplace (or Borel) transform on both sides, and 
transferring the continuum contribution to the QCD side, the moment sum rule
and its ratio read:
\bea {\cal F}_i(\tau)&\equiv& \lambda_i^2e^{-M_i^2\tau}=\int_{4m_c^2}^{t_c}ds~
e^{-s\tau}~\rho_i(s)\; \nnb\\
{\cal R}_i(\tau)&\equiv& -{d\over d\tau}{\log {\cal F}_i(\tau)}\simeq M_i^2~
\lb{sr1} 
\eea
where $\tau\equiv 1/M^2$ is the sum rule variable with $M$ being the 
inverse-Laplace (or Borel) mass. 
In the following, we shall work with the DRSR \cite{DRSR}:
\bea 
r_{T_{cc}/X}&\equiv& \sqrt{{\cal R}_{T_{cc}}\over {\cal R}_X}\simeq {M_{T_{cc}}
\over M_X}.
\lb{drsr} \eea

\subsection{QCD expression of the spectral functions}
The QCD expressions of the spectral densities of the two-point correlator 
associated to the current in Eq. (\ref{currx}) are given
in ref.~\cite{x24}. Up to dimension-six condensates
the expressions associated to the current in Eq. (\ref{currt}), in the
structure $g_{\mu\nu}$ are: 
\bea
\rho(s)&=&\rho^{pert}(s)+\rh^{\qq}(s)+\nnb\\
&&\rh^{\lag G^2\rag}
(s)+\rh^{mix}(s)+\rh^{\qq^2}(s)\;,
\lb{rhoeq}
\eea
with

\beqa
\label{eq:pert}
&&\rho^{pert}(s)={1\over3. 2^{13} \pi^6}\int\limits_{\almi}^{\alma}
{d\al\over\alpha^3}
\int\limits_{\bemi}^{1-\al}{d\be\over\be^3}(1-\al-\be)\times\nnb\\
&&\times\left[(\al+\be)m_c^2-\al\be s\right]^3\bigg[2m_c^2(13\al^2+\nnb\\
&&+13\al\be+7\al+5)-15\al\be s(1+\al+\be)\bigg],
\nn\\
&&\rho^{\qq}(s)=-{m_c\qq\over 2^{6}\pi^4}\int\limits_{\almi}^{\alma}
{d\al\over\al}
\int\limits_{\bemi}^{1-\al}{d\be\over\be^2}(1+\al+\be)\times\nnb\\
&&\times\left[(\al+\be)m_c^2-
\al\be s\right]^2,\nn\\
&&\rho^{\la G^2\ra}(s)={\Gd\over2^{13}.3^2\pi^6}\int\limits_{\almi}^{\alma} {d\al
\over\al^2}\int\limits_{\bemi}^{1-\al}{d\be\over\be^3}\bigg\{12\al\be(5\al+\nn\\
&&+5\be-3)\left[(\al+\be)m_c^2-\al\be s\right]^2+m_c^4(1-\al-\be)^2)\times\nn\\
&&\times\al^2(5+\al+\be)
+m_c^2(1-\al-\be)\Big{[}9\al^2(2+\nn\\
&&+3\al+4\be)+\al\be(2+2\al+11\be)+\nn\\
&&+15\al-2\be\Big{]}\left[(\al+\be)m_c^2-\al\be s\right]
\bigg\}, \nn\\
&&\rho^{mix}(s)=-{m_c\mix\over3. 2^{8}\pi^4}\bigg\{7\int\limits_{\almi}^{\alma}
{d\al\over\al(1-\al)}\left[m_c^2+\right.\nn\\
&&\left.-\al(1-\al)s\right]-\int\limits_{\almi}^{\alma}
{d\al\over\al}\int\limits_{\bemi}^{1-\al}{d\be\over\be^2}\left[12\al^2+\right.\nn\\
&&\left.+17\al\be-\be
\right]\left[(\al+\be)m_c^2-\al\be s\right]\bigg\},\nnb\\
&&\rho^{\qq^2}(s)={\rho\qq^2\over3. 2^6\pi^2}\int\limits_{\almi}^{\alma}d\al
\left[13m_c^2-5\al(1-\al)s\right],
\enqa
where: $m_c,~\lag
g^2G^2\rag,~\qq, ~\mix$ are respectively the charm quark mass, gluon condensate, 
light quark and mixed condensates; 
 $\rho$ indicates the violation of the four-quark vacuum 
saturation.
The integration limits are given by:
 \bea
 \almi&=&{1\over 2}({1-v}),~~~~ \alma={1\over 2}({1+v})\nnb\\
 \bemi&=&{\al m_c^2/( s\al-m_c^2)}
 \label{eq:limit}
\eea
where $v$ is the $c$-quark velocity:
\beq
v\equiv  \sqrt{1-4m_c^2/s}~.
\eeq
\subsection{$T_{cc}/X$ ratio of masses}
In the following, we shall extract the mass ratio $T_{cc}/X$ using the DRSR in Eq. (\ref{drsr}).
For the numerical analysis we shall  introduce the renormalization group invariant  
quantities 
$\hat m_s$ and $\hat\mu_q$ \cite{FNR,TARRACH}:
\bea
{\bar m_s}(\tau)&=&{\hat m_s \over \ga-\log{ \sqrt{\tau}\Lambda}\dr^{-2/{
\beta_1}}}\nnb\\
{\la\bar qq\ra}(\tau)&=&-{\hat \mu_q^3 \ga-\log{ \sqrt{\tau}\Lambda}\dr^{-2/{
\beta_1}}}\nnb\\
{\la\bar qg\sigma.Gq\ra}(\tau)&=&-m_0^2{\hat \mu_q^3 \ga-\log{ \sqrt{\tau}
\Lambda}\dr^{-1/{3\beta_1}}}~,
\eea
where $\beta_1=-(1/2)(11-2n/3)$ is the first coefficient of the $\beta$ function 
for $n$ flavours. We have used the quark mass and condensate anomalous dimensions 
reported in \cite{SNB}.  We shall use the QCD parameters in 
Table \ref{tab:param}. At the scale where we shall work, and using the parameters 
in Table \ref{tab:param}, we deduce: $\rho=2.1\pm 0.2$,
which controls the deviation from the factorization of the four-quark condensates. 
We shall not include the $1/q^2$ term discussed in \cite{CNZ,ZAK},which is 
consistent with the LO approximation 
used here as the latter has been motivated by a phenomenological parametrization  
of the larger order terms of the QCD series.

{\scriptsize
\begin{table}[hbt]
\setlength{\tabcolsep}{0.2pc}
 \caption{
QCD input parameters. For the heavy quark masses, we use 
the range spanned
 by the running $\overline{MS}$ mass $\overline{m}_Q(M_Q)$  and the on-shell mass 
from QCD (spectral) sum rules compiled in page
 602,603 of the book in \cite{SNB}. The values of $\Lambda$ and $\hat\mu_q$ have 
been obtained from $\alpha_s(M_\tau)=0.325(8)$ \cite{SNTAU} and from the running 
masses:  $(\overline{m}_u+\overline{m}_d)(2)=7.9(3)$ MeV \cite{SNmass}. The 
original errors have been multiplied by 2 for a conservative estimate of the 
errors.     }
    {\small
\begin{tabular}{lll}
&\\
\hline
Parameters&Values& Ref.    \\
\hline
$\Lambda(n_f=4)$& $(324\pm 15)$ MeV &\cite{SNTAU,PDG}\\
$\hat \mu_q$&$(263\pm 7)$ MeV&\cite{SNB,SNmass}\\
$\hat m_s$&$(0.114\pm0.021)$ GeV &\cite{SNB,SNmass,PDG}\\
$m_c$&$(1.23\sim1.47)$ GeV &\cite{SNB,SNmass,SNHmass,PDG,SNH10,IOFFE}\\
$m_b$&$(4.2\sim4.7)$ GeV &\cite{SNB,SNmass,SNHmass,PDG,SNH10}\\
$m_0^2$&$(0.8 \pm 0.2)$ GeV$^2$&\cite{JAMI2,HEID,SNhl}\\
$\la\alpha_s G^2\ra$& $(6\pm 2)\times 10^{-2}$ GeV$^4$&
\cite{SNTAU,LNT,SNI,fesr,YNDU,SNHeavy,BELL,SNH10,SNG}\\
$\rho\alpha_s\la \bar dd\ra^2$& $(4.5\pm 0.3)\times 10^{-4}$ GeV$^6$&
\cite{SNTAU,LNT,JAMI2}\\
\hline
\end{tabular}
}
\label{tab:param}
\end{table}
}

Using QCD (spectral) sum rules, one can usually estimate the mass of the $X$-meson, from the ratio $
{\cal R}_{X}$ in Eq. (\ref{sr1}), which is related to the spectral densities 
obtained from the current (\ref{currx}). A tetraquark current for the $X(3872)$ 
was used in ref.~\cite{mnnr}. At the sum rule
stability point and using a slightly different (though consistent) set of QCD 
parameters than in Table \ref{tab:param},  one obtains, with a good accuracy,
for $m_c=1.26$ GeV \cite{mnnr} \footnote{The use of $m_c=1.47$ GeV increases the central
value by about $(160\sim 200)$ MeV.}:
\beq
M_X\simeq \sqrt{{\cal R}_{X}}= (3925\pm 127)~{\rm MeV}~,
\lb{m2}
\enq
and the correlated continuum threshold value fixed simultaneously by the Laplace 
and finite energy sum rules 
(FESR) sum rules:
\beq
\sqrt{t_c}\simeq (4.15\pm 0.03)~{\rm GeV}~.
\lb{tc3}
\enq
In ref.~\cite{drx} it was obtained that the DRSR for the tetraquark
current and for the molecular current is:
\beq
r_{mol/3}=\sqrt{{\cal R}_{mol}\over{\cal R}_{3}}\simeq 1.00~,
\lb{drmol}
\enq
with  a negligible error. Therefore, the result in Eq.~(\ref{tc3}) is the 
same for the current
in Eq.~(\ref{currx}). Although the uncertainty in Eq.~(\ref{m2}) is still large,
considering the fact that this result was obtained in a Borel region
where there is pole dominance and OPE convergence, one can say that the
QCD sum rules supports the existence of such a state and that the
value obtained for $M_X$ is in reasonable agreement  with the experimental 
candidate \cite{PDG}:
\beq
M_X\vert_{exp}= (3872.2\pm 0.8)~{\rm MeV}~.
\lb{Xexp}
\enq

\begin{figure}[hbt] 
\begin{center}
\centerline{\includegraphics[height=60mm]{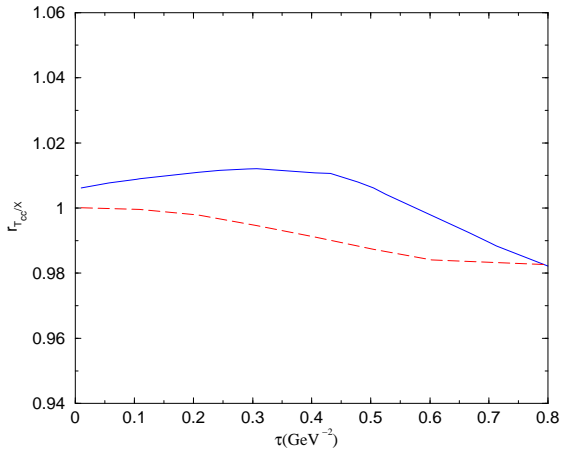}}
\caption{
The double ratio $r_{T_{cc}/X}$ defined in 
Eq. (\ref{drsr}) as a function of 
$\tau$ for $\sqrt{t_c}
=4.15~\GeV$ and for two values of $m_c=1.23$ (solid line) and 1.47 GeV (dashed line).}
\label{fig1} 
\end{center}
\end{figure} 

\begin{figure}[h] 
\centerline{\includegraphics[height=60mm]{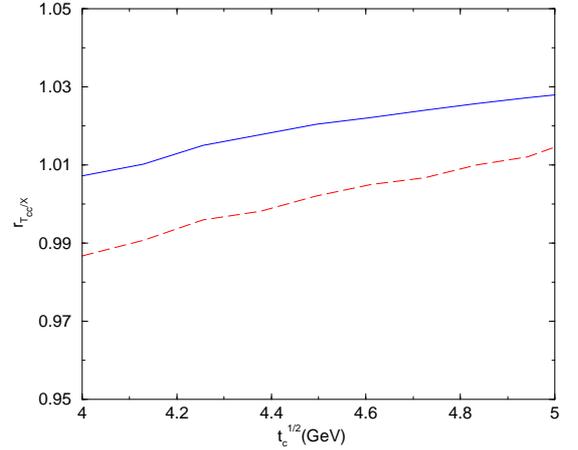}}
\caption{
The double ratio $r_{T_{cc}/X}$  as a function of 
$t_c$ for $\tau=0.4~\GeV^{-2}$ and for two values of $m_c=1.23$ (solid line) 
and 1.47 GeV (dashed line).}
\label{fig2} 
\end{figure}

We now study the DRSR of the 
$T_{cc}/X$ defined in Eq.~(\ref{drsr}). In Fig. \ref{fig1}, we show the $\tau$-dependence
of the ratio for $\sqrt{t_c}=4.15~\GeV$ and for two values of $m_c=1.23~\GeV$ and 1.47 GeV.
>From Fig. \ref{fig1} one can see that there is a  $\tau$-stability around $\tau\simeq 0.4~
\GeV^{-2}$ and for this value of $\tau$, we get:
\beq
r_{T_{cc}/X}=1.00\pm 0.01~.
\label{eq:ratiolam}
\eeq
In Fig. \ref{fig2}, we show the $t_c$-dependence
of the ratio for $\tau=0.4~\GeV^{-2}$ and for two values of $m_c=1.23~\GeV$ and 1.47 GeV.
>From this figure one can see that the ratio increases with $t_c$. However, considering
the large range of $t_c$ presented in the figure, the ratio does not differ more than
3\% from 1. 

Our analysis has shown that the  $D\bar{D}^*+c.c.$ and $DD^*$ currents lead to
the same mass predictions within the accuracy of the approach. The accuracy
of the DRSR is bigger than the normal QCDSR because the DRSR are less 
sensitive to the exact value and definition of the heavy quark mass and to 
the QCD continuum contributions . As mentioned before, this accuracy is 
reached due to partial cancellations of the systematics of the method and of 
the QCD corrections in the DRSR.  
Therefore, if the observed $X(3872)$ is a molecular $D\bar{D}^*+c.c.$ state
its molecular cousin $DD^*$ should also be a bound state. Its mass can be 
obtained by using the experimental mass for the  $X(3872)$ in 
Eq.~(\ref{eq:ratiolam}):
\bea 
M_{T_{cc}}=(3872.2\pm39.5)~\MeV
\lb{mtcc} \eea

\subsection{$T_{bb}/X_b$ ratio of masses}
Using the same interpolating field in Eqs.~(\ref{currx}) and (\ref{currt}) 
with the charm quark replaced by the bottom one, we can analyse the
DRSR:
\bea 
r_{T_{bb}/X_b}&\equiv& \sqrt{{\cal R}_{T_{bb}}\over {\cal R}_{X_b}}\simeq 
{M_{T_{bb}}\over M_{X_b}}~.
\lb{drb} \eea

\begin{figure}[hbt] 
\begin{center}
\centerline{\includegraphics[height=60mm]{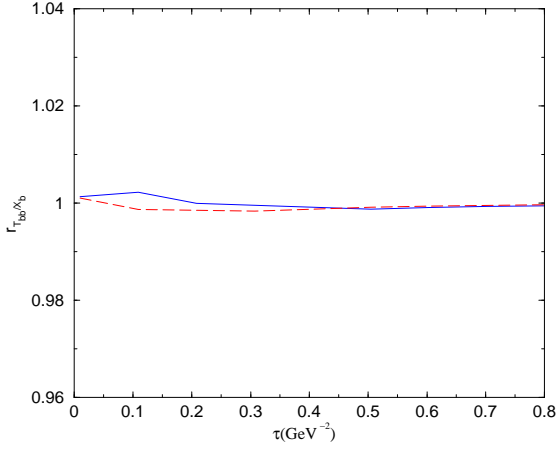}}
\caption{
Same as Fig.~\ref{fig1} for $r_{T_{bb}/X_b}$ for $\sqrt{t_c}
=10.5~\GeV$ and for two values of $m_b=4.2$ (solid line) and 4.7 GeV (dashed line).} 
\label{fig3} 
\end{center}
\end{figure} 

\begin{figure}[hbt] 
\begin{center}
\centerline{\includegraphics[height=60mm]{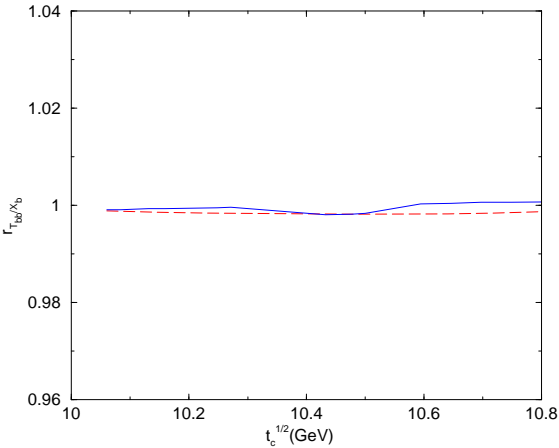}}
\caption{
Same as Fig.~\ref{fig2} for $r_{T_{bb}/X_b}$ for $\tau
=0.2~\GeV^{-2}$ and for two values of $m_b=4.2$ (solid line) and 4.7 GeV (dashed line).} 
\label{fig4} 
\end{center}
\end{figure}

In Fig. \ref{fig3}, we show the $\tau$-dependence
of the ratio in Eq.~(\ref{drb}) for $\sqrt{t_c}=10.5~\GeV$ and for two values
of $m_b$. 
>From this figure one can see the ratio is very stable. The same happens for
the dependence of this ratio with $t_c$, as can be seen by Fig.~\ref{fig4}. We get:
\beq
r_{T_{bb}/X_b}=1.00~.
\label{eq:ratiolam2}
\eeq
Therefore, we can predict the degeneracy between the masses of the $T_{bb}$ and of the $X_b$ given in Eq.~(\ref{massXb}).

\section{\boldmath Revisiting the determination of the $X_b$ mass }
\begin{figure}[H] 
\centerline{\includegraphics[height=50mm]{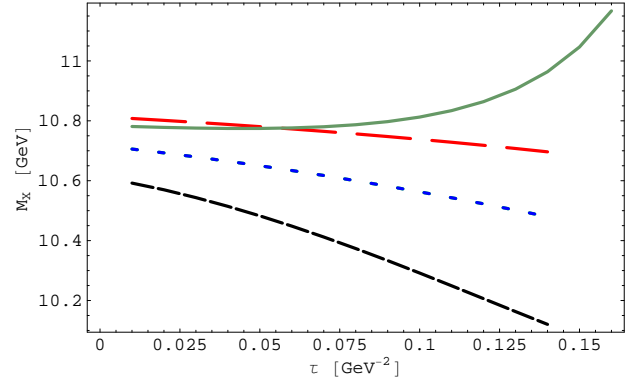}}
\caption{
$M_{X_b}$  in GeV as a function of 
$\tau$ in GeV$^{-2}$ from Laplace sum rule for $m_b=4.7$ GeV and $\sqrt{t_c}=11$ GeV: long dashed (red) : (1)= perturbative (Pert) contribution; small dashed (blue): (2)= Pert+$\la \bar dd\ra +\la \alpha_s G^2\ra$ contributions (the one of the gluon condensate is relatively negligible); continuous (olive): (3)=(2)+ mixed condensate $\la g\bar d G d\ra$; medium dashed (black): (4)=(3)+$\la \bar dd\ra^2$.}
\label{fig:mxtau} 
\end{figure} 
\begin{figure}[h] 
\centerline{\includegraphics[height=50mm]{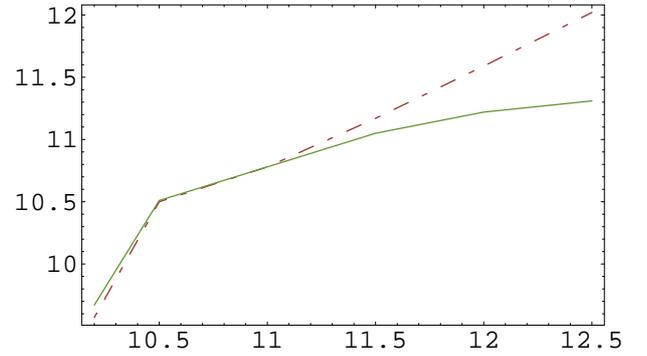}}
\caption{
$M_{X_b}$  in GeV as a function of 
$\sqrt{t_c}$ in GeV from FESR (dashed curve) and  the value at the $\tau$-stability from Laplace sum rule for $m_b=4.7$ GeV. The OPE has been trunctaed at $D=5$.}
\label{fig:mx} 
\end{figure} 
The $X_b$ was studied in ref.~\cite{mnnr}. At the sum rule
stability point and using the perturbative $\overline{MS}$-mass 
$m_b(m_b)=4.24~\GeV$, they get:
\beq
\lb{massXb}
10.06~\GeV \leq M_{X_b}\leq   10.50~\GeV~,
\enq
for  $10.2~\GeV \leq \sqrt{t_c}\leq10.8~\GeV $, while combining the Laplace sum rule with FESR,
they obtain a slightly lower but more precise value :
\beq
M_{X_b}=(10.14\pm 0.10)~\GeV~.
\lb{eq:massXb2}
\eeq
We complete the previous analysis by using here the value of the on-shell mass $m_b= 4.7$ GeV due to the
ambiguous definition of the quark mass used as we work to leading order of radiative corrections. For a close comparison with the analysis in \cite{mnnr}, we shall work with the two-point function associated
to the four-quark current \footnote{The result using the $D\bar D^*$molecule current would be the same
as we have shown in \cite{drx} that the masses obtained from the four-quark and molecule currents are
degenerate.}.
We notice that the contribution of the $D=6$ condensate $\la \bar qq\ra^2$ destabilizes the result [medium dashed (black) curve] \footnote{We have neglected the contribution of the triple gluon condensate $\la g^3f_{abc}G^3\ra$ due to the $(1/16\pi^2)^2$ loop factor suppression compared to $\la \bar qq\ra^2$.}, which is restored if one adds the $D= 8$ condensate given in \cite{mnnr}. However, we refrain to add such a term due to the eventual uncertainties for controlling the high-dimension condensate contributions (violation of factorization, complete $D= 8$ contributions) and find safer to limit the analysis to the $D=5$ contribution like in Ref. \cite{mnnr}. In this way, the ratio of sum rules present $\tau$-stability at about 0.1 GeV$^{-2}$ (continuous curve in  Fig \ref{fig:mxtau}). We show in Fig. \ref{fig:mx} the $t_c$-behaviour of $M_{X_b}$ versus the
continuum threshold $t_c$, where a common solution is obtained in units of GeV:
\beq
M_{X_b}= 10.50\sim 10.78~~~~{\rm for}~~~\sqrt{t_c}=10.5\sim 11.0~,
\eeq
which combined with the result in Eq. (\ref{eq:massXb2}) leads to the conservative range of values:
\beq
M_{X_b}= 10.14\sim 10.78~~~~{\rm for}~~~\sqrt{t_c}\approx M_{X'_b}=10.5\sim 11.0~,
\lb{eq:massxb}
\eeq
where one can notice the relatively small mass-difference between $\sqrt{t_c}$ and $M_{X_b}$ eventually signaling the nearby location of the radial excitations as mentioned in \cite{mnnr}.\\
Very recently the Belle Collaboration studied the $\Upsilon(5S)\to\Upsilon
(nS)\pi^\pm$ and $\Upsilon(5S)\to h_b(mP)\pi^\pm$ ($n=1,2,3$ and $m=1,2$)
decay processes looking for  resonant substructures. 
They  found two narrow states in units of MeV: 
\beq
M_{Z_{b}}=10610~~~~~~~{\rm  and}~~~~~~~M_{Z'_{b}}=10650~,
\lb{eq:massexp}
\eeq
with a hadronic width in units of MeV: 
\beq
\Gamma_{Z_{b}}= 15.6 \pm 2.5~~~~ {\rm and}~~~~\Gamma_{Z'_{b}}= 14.4 \pm 3.2~,
\eeq 
respectively  \cite{bellezb}. 
The analysis of the $Z_b$ states decay in the channel  
$Z_b^+  \rightarrow \Upsilon(2S) \pi^+ $  favors the $J^P = 1^+$ assignment, 
which is the same as the one of the $X_b$, although $X_b$ has positive charge
conjugation and the neutral partiner of $Z_b$ should have negative
charge conjugation.  \\

Considering the errors in Eq. (\ref{eq:massxb}) and the small mass 
difference between  $X_b$ and $\sqrt{t_c}$, it is difficult to identify
the two observed $Z_b$ states with the $X_b$.

\section{\boldmath $SU(3)$ Mass-splittings }
We extend the previous analysis to study the ratio between the strange $T^s_{cc}$
($(cc\bar s\bar s)$) and non-strange $T_{cc}$ states. \\
The QCD expression for the spectral function proportional to $m_s$ 
for the current in Eq.~(\ref{currt}) is:

\bea
&&\rho^{m_s}(s)={m_s \over 2^8 \pi^4} \int\limits_{\almi}^{\alma}
{d\al\over\al} 
\bigg\{ {\ss}{7[m_{c}^2-\al(1-\al)s]^2 \over (1-\al)}\nn\\
&&+ \int\limits_{\bemi}^{1-\al}{d\be\over\be}\kab\bigg[\ss\bigg(  
m_{c}^{2}(21+\al+\be)\nn\\
&&-6\kab\bigg)-{m_c\over2\pi^2\al\be^2}(3+\al+\be)\nn\\
&&\times(1-\al-\be)\kab^2\bigg]\bigg\}.
\eea

We start by studying the DRSR:
\bea 
r_{T^s_{cc}/T_{cc}}&\equiv& \sqrt{{\cal R}_{T^s_{cc}}\over {\cal R}_{T_{cc}}}
\simeq {M_{T^s_{cc}}\over M_{T_{cc}}}.
\lb{drsu3} \eea
\begin{figure}[hbt] 
\begin{center}
\centerline{\includegraphics[height=60mm]{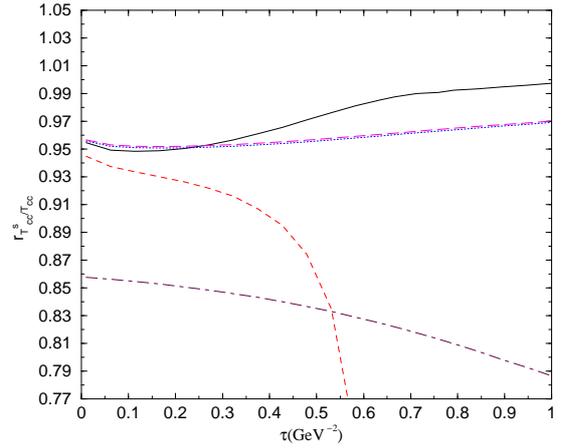}}
\caption{
Different QCD contributions to the DRSR $r_{T^s_{cc}/T_{cc}}$ defined in 
Eq. (\ref{drsu3})  for $\sqrt{t_c}
=4.15~\GeV$ and $m_c=1.23$ GeV:  (1)=Pert+$m_s$: dot-dashed (maroon); (2)=(1)
+$\la\bar qq\ra$: long-dashed (magenta); (3)=(2)+$\la \alpha_s G^2
\ra$: dotted (blue); (4)=(3)+$\la g\bar q G q\ra$: dashed (red);  (5)=(4)
+$\la\bar qq\ra^2$: continuous (black).}
\label{fig:drsrope} 
\end{center}
\end{figure} 
\begin{figure}[hbt] 
\begin{center}
\centerline{\includegraphics[height=60mm]{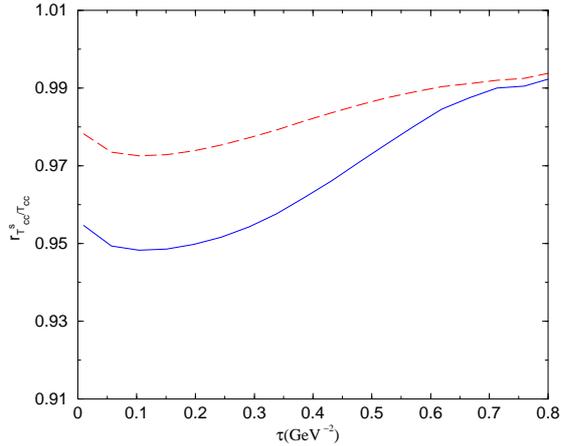}}
\caption{
Same as Fig.~\ref{fig1} for $r_{T^s_{cc}/T_{cc}}$ defined in 
Eq. (\ref{drsu3})  for $\sqrt{t_c}
=4.15~\GeV$ and for two values of $m_c=1.23$ (solid line) and 1.47 GeV (dashed line).}
\label{fig5} 
\end{center}
\end{figure} 
We study the convergence of the DRSR versus $\tau$ in Fig \ref{fig:drsrope}, where
the show the strength of the different contributions in the OPE. We note that we have $\tau$-stability 
around $\tau=0.2$ GeV$^{-2}$, while the OPE breaks down for $\tau\geq 0.5$ GeV$^{-2}$.

\begin{figure}[hbt] 
\begin{center}
\centerline{\includegraphics[height=60mm]{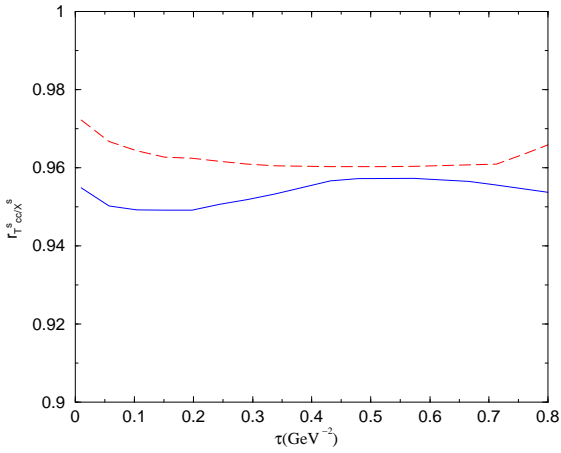}}
\caption{
Same as Fig.~\ref{fig1} for $r_{T^s_{cc}/X^s}$. The solid and dashed lines
are for $m_c=1.23$  and 1.47 GeV respectively.}
\label{fig6} 
\end{center}
\end{figure}

In Fig. \ref{fig5}, we show the $\tau$-dependence
of the ratio in Eq.~(\ref{drsu3}), for $\sqrt{t_c}=4.15~\GeV$ and for two values of
$m_c$. From this Figure one can deduce around the  $\tau$-stability:
\beq
r_{T^s_{cc}/T_{cc}}=0.95\sim 0.98~,
\label{eq:rasu3}
\eeq
which gives a smaller mass for $T^s_{cc}$ than for $T_{cc}$. This result is similar
to the result obtained for $X^s$ in ref.~\cite{mnnr}. However, in this case, the
decrease in the mass is even bigger than the obtained for $X^s$: 
$r_{X^s/X}=0.984\pm 0.009$. This result is consitent with what is obtained from the
DRSR:
\bea 
r_{T^s_{cc}/X^s}&\equiv& \sqrt{{\cal R}_{T^s_{cc}}\over {\cal R}_{X^s}}
\simeq {M_{T^s_{cc}}\over M_{X^s}},
\lb{drs} 
\eea
as can be seen by Fig.~\ref{fig6}.

\begin{figure}[hbt] 
\begin{center}
\centerline{\includegraphics[height=60mm]{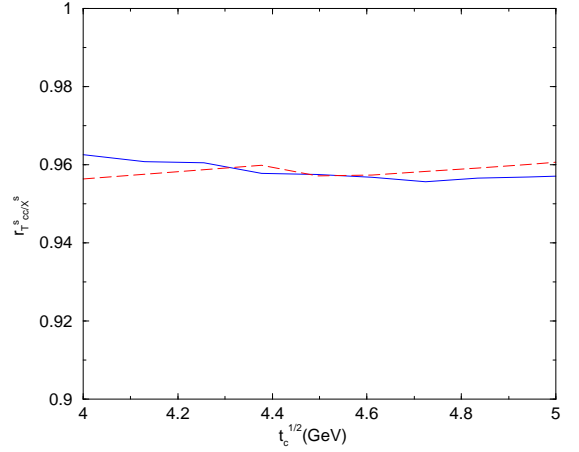}}
\caption{
Same as Fig.~\ref{fig2} for $r_{T^s_{cc}/X^s}$. The solid and dashed lines
are for $m_c=1.23$  and 1.47 GeV respectively.}
\label{fig7} 
\end{center}
\end{figure}

In Fig.~\ref{fig6} we have used $t_c=4.15~\GeV$. However, the result is very stable 
as a function of $t_c$, as can be seen in Fig.~\ref{fig7}.

Since for the $T^s_{cc}$ state, considered as a $D_sD_s^*$ molecule, there is no 
allowed pion exchange, one can not conclude, from the analysis above, that the 
$T^s_{cc}$ should be more deeply bound than the $T_{cc}$. On the contrary, if the pion
exchange is important for binding the two mesons, the $T^s_{cc}$ may not be bound.

\section{Conclusion}
In conclusion, 
we have studied the mass of the $T_{cc}$ using double ratios of sum rules (DRSR), 
which are more accurate
than the usual simple ratios used in the literature. We found that the molecular 
currents
 $D\bar{D}^*+c.c.$ and $DD^*$ lead to (almost) the same mass predictions 
within the accuracy of the method. Since the pion exchange interaction between
these mesons is exactly the same, we conclude that if the observed $X(3872)$ meson
is a $D\bar{D}^*+c.c.$ molecule, then the $DD^*$ molecule should also exist with
approximately the same mass. A recent estimate of the production rate indicates that 
these states could be seen in LHC experiments \cite{Yuqi:2011gm}.

We have also studied the double ratio $r_{T_{bb}/X_b}$ using  molecular currents
$\bar{B}\bar{B}^*$ and $B\bar{B}^*+c.c.$ for $T_{bb}$ and $X_b$ respectively. In
this case the degeneracy between the two masses is even better than in the charm case.
Therefore, we also conclude for the bottom case  that  if a molecular state
$B\bar{B}^*+c.c.$ exist, then the $\bar{B}\bar{B}^*$ molecule should also exist with
the same mass.

\vskip1.5cm

\subsection*{Acknowledgment}

\noindent
This work has been partly supported 
by the CNRS-FAPESP program,  by  CNPq-Brazil and by the CNRS-IN2P3 
within the project Non-perturbative QCD and Hadron Physics. We thank R.M. Albuquerque
for checking some of the QCD expressions of the two-point correlator.

\end{document}